\documentclass[groupedaddress,showpacs,showkeys,amssymb,nofootinbib,aps]{revtex4}      
\usepackage[pdftex]{graphicx}
\usepackage{amsmath} 
\usepackage{epsf}                                                                                           
\usepackage{color}                   
\usepackage{verbatim}                                                                         

\newcommand{\be}{\begin{equation}}

\newcommand{\ee}{\end{equation}}

\begin{document}                                                                                              
                                                                                                              
\title {On Restricting to One Loop Order the Radiative Effects in
  Quantum Gravity}






\author{F. T. Brandt and J. Frenkel} 
\email{fbrandt@usp.br, jfrenkel@if.usp.br} 
\affiliation{Instituto de F\'{\i}sica, Universidade de S\~ao Paulo, S\~ao Paulo, SP 05508-090, Brazil}
\author{D. G. C. McKeon}
\email{dgmckeo2@uwo.ca}
\affiliation{
Department of Applied Mathematics, The University of Western Ontario, London, ON N6A 5B7, Canada}
\affiliation{Department of Mathematics and Computer Science, Algoma University,
Sault St.Marie, ON P6A 2G4, Canada}

\date{\today}
                                                                                                              
\begin{abstract}
The dimensionful nature of the coupling in the Einstein-Hilbert action in four dimensions implies that the theory
is non-renormalizable; explicit calculation shows that beginning at
two loop order, divergences arise that cannot be removed by
renormalization without introducing new terms in the classical
action. It has been shown that, by use of a Lagrange multiplier field
to ensure that the classical equation of motion is satisfied in the
path integral, radiative effects can be restricted to one loop
order. We show that by use of such Lagrange multiplier fields, the
Einstein-Hilbert action can be quantized without the occurrence of
non-renormalizable divergences.  We then apply this mechanism to a
model in which there is in addition to the Einstein-Hilbert action, a
fully covariant action for a self-interacting scalar field coupled to
the metric. It proves possible to restrict loop diagrams involving
internal lines involving the metric to one-loop order; diagrams in
which the scalar field propagates occur at arbitrary high order in
the loop expansion. This model also can be shown to be
renormalizable. Incorporating spinor and vector fields in the same way
as scalar fields is feasible, and so 
a fully covariant Standard Model 
with a dynamical metric field
can also be shown to be renormalizable
\end{abstract}                                                                                                

                                                               
\maketitle                     


\section{Introduction}

Removing divergences arising from loop momentum integrals is a
particularly acute problem in quantum gravity due to the dimensionful
nature of the coupling. The divergences occurring at one-loop order
when using the Einstein-Hilbert action can be removed by a field
redefinition on account of 
the divergences vanishing if the equations of motion are
satisfied if
the Gauss-Bonnet identity 
is used
\cite{1,2}, but
once the metric interacts with a scalar \cite{1}, vector \cite{3} or
spinor \cite{4} field this is no longer possible even at one-loop
order \footnote{By using analytic continuation, divergences can be avoided completely \cite{5}.}.  
Not even the Einstein-Hilbert action by itself is renormalizable, in the power-counting sense, 
beyond one-loop order \cite{6,7}.
It is well known \cite{D,W}, that quantum gravity
based on the Einstein-Hilbert action
is renormalizable 
if
there is a counter-term available to cancel every ultraviolet divergence.
However, this procedure requires an infinite number of counter-terms, which
lessens the predictive power of the theory. 

A way has been found to eliminate all radiative effects beyond one-loop order in the loop expansion.  This has been illustrated in Yang-Mills theory \cite{8} and the Proca model \cite{9}.
By  using a Lagrange multiplier field to impose the condition that
when evaluating the quantum path integral, only field configurations
that satisfy the classical equations of motion contribute  
and one no longer encounters radiative effects beyond one loop.
The tree-level diagrams are reproduced and the one-loop contribution is twice that of the usual one-loop diagrams that occur without this Lagrange multiplier field; all contributions beyond one-loop order are absent.  The problem of showing renormalizability is thus greatly simplified as only one-loop effects need to be considered.
This procedure is also consistent with unitarity.

We first show how this approach using a Lagrange multiplier field can be used in conjunction
with the Einstein-Hilbert action alone. In this case, upon using the Gauss-Bonnet theorem,
the divergences arising from one-loop effects can be removed by a 
shift
of the
Lagrange multiplier field. Next, we add to this action, the fully covariant action of a
self interacting scalar field. This results in diagrams of arbitrary high order in
the loop expansion, but we still find that propagators involving the metric field only contribute
to one-loop diagrams. The model remains renormalizable, even when considering these higher loop
diagrams involving internal scalar field lines. It is possible to couple the metric not only to
a scalar field, but also to all fields contributing to the Standard Model, again in a way
that leaves the theory renormalizable.

\section{Use of a Lagrange Multiplier}

In general, an action
\begin{equation}\label{eq1}
S[\phi_i] = \int dx \left( \mathcal{L} [\phi_i (x)] + j_i \phi_i\right)
\end{equation}
can be considered in conjunction with the path integral \cite{8,9}
\begin{align}\label{eq2}
Z^2_j = \lim\limits_{\eta \rightarrow \infty} \int {\cal D}\phi_{i} & {\cal D}\lambda_{i} \exp i \Bigg[
\left(\frac{1+\eta}{2}\right) S[\phi_{+_{i}}]\\
&+ \left(\frac{1-\eta}{2}\right) S[\phi_{-_{i}}]\Bigg]\nonumber\\
&\hspace{2cm} \left( \phi_{\pm_{i}} \equiv \phi_i \pm \frac{1}{\eta}\lambda_i\right)\nonumber
\end{align}
\begin{align}\label{eq3}
&= \int {\cal D}\phi_i  {\cal D}\lambda_i \exp i \int dx \Bigg[ \mathcal{L} [\phi_k] + \lambda_i 
\frac{\delta\mathcal{L}[\phi_k]}{\delta\phi_i}\nonumber \\
&\hspace{2cm}+ j_i\left(\phi_i + \lambda_i\right)\Bigg].
\end{align}
Integration over the Lagrange multiplier field $\lambda_i$ leads to
\begin{equation}\label{eq4}
Z^2_j = \int {\cal D}\phi_i\; \delta \left[ \frac{\delta\mathcal{L}}{\delta\phi_i} + j_i\right] \exp i \int dx \left[ \mathcal{L} [\phi_i] + j_i\phi_i\right].
\end{equation}
The functional analogue to
\begin{equation}\label{eq5}
\int dx \; \delta \left( f(x)\right) g(x) = \sum_i \frac{g(x_i)}{|f^\prime(x_i)|} \qquad \left(f(x_i) = 0\right)
\end{equation}
reduces Eq. \eqref{eq4} to
\begin{equation}\label{eq6}
Z^2_j = \sum_i \exp i \int dx \left[ \mathcal{L} [\overline{\phi}_i] + j_i\overline{\phi}_i\right]/
\det \left( \frac{\delta^2\mathcal{L}[\overline{\phi}_k]}{\delta\phi_i\delta\phi_j}\right),
\end{equation}
where $\frac{\delta\mathcal{L}[\overline{\phi}_k]}{\delta\phi_i} + j_i = 0$ defines $\overline{\phi}_i$.  In Eq. \eqref{eq6}, the exponential is the sum of all tree-level diagrams while the functional determinant is the square of the usual one-loop contribution when there is no Lagrange multiplier $\lambda_i$ present.

A diagrammatic approach to the path integral of Eq. \eqref{eq3} uses the expansion
\begin{equation}\label{eq7}
\mathcal{L} [\phi_i] = \frac{1}{2!} a_{ij} \phi_i\phi_j + \frac{1}{3!} a_{ijk} \phi_i\phi_j \phi_k + \ldots .
\end{equation}
The bilinear $\frac{1}{2} a_{ij}(\phi_i\phi_j + 2\phi_i \lambda_j)$ leads to the propagators $<\phi_i\phi_j> = 0$, $<\phi_i\lambda_j> = a_{ij}^{-1} = -<\lambda_i\lambda_j>$ since 
$\begin{pmatrix} a & a\\ a&0\end{pmatrix}^{-1} = 
\begin{pmatrix} 0 & a^{-1}\\ a^{-1}& -a^{-1}\end{pmatrix}.$  As
$<\phi_i \phi_j> = 0$ and since all vertices are at most linear in
$\lambda_i$, the only Feynman diagrams that can contribute have mixed
propagators  $<\phi_i \lambda_j>$ with only the fields $\phi_i$ on
external legs.  A combinatorial analysis shows that these diagrams are
twice the corresponding one-loop diagrams that come from
Eq. \eqref{eq1} \cite{8}.

If there is an infinitesimal gauge symmetry
\begin{equation}\label{eq8}
\phi_i \rightarrow \phi_i^\prime = \phi_i + H_{ij}(\phi_k)\xi_j
\end{equation}
in Eq. \eqref{eq1}, then $a_{ij}$ in Eq. \eqref{eq7} cannot be
inverted. In this case Eq. \eqref{eq1} requires the addition of a gauge fixing Lagrangian
\begin{equation}\label{eq9}
\mathcal{L}_{gf} = - \frac{1}{2\alpha} \left(F_{ij} \phi_j\right)^2
\end{equation}
and a ghost Lagrangian
\begin{equation}\label{eq10}
\mathcal{L}_{ghost} = \overline{c}_i F_{ij} H_{jk} c_k
\end{equation}
when using the path integral \cite{10,11}.

The invariance of Eq. \eqref{eq8} means that
\begin{eqnarray}\label{eq11}
  \int dx {\cal L}[\phi^\prime_i] &=& \int dx {\cal L}[\phi_i]\nonumber \\
  &=& \int dx\left( {\cal L}[\phi_k]+H_{ij}[\phi_k]\xi_j
  \frac{\delta{{\cal L}[\phi_k]}}{\delta\phi_i}\right)
\end{eqnarray}
and consequently $\int d x \lambda_i\frac{{\delta\cal L}[\phi_k]}{\delta\phi_i}$ is invariant
under the transformation
\be\label{eq12}
\lambda_i\rightarrow\lambda_i+H_{ij}[\phi_k] \zeta_k.
\ee
Furthermore, since by Eq. \eqref{eq11}
\begin{eqnarray}\label{eq13}
&&  \int dx \left({\cal L}[\phi_k]+
  \lambda_i\frac{{\delta\cal L}[\phi_k]}{\delta\phi_i}\right)\nonumber \\
&=& 
\int dx \left({\cal L}[\phi^\prime_k]+
\lambda_i\frac{\delta\phi^\prime_j}{\delta\phi_i}
\frac{{\delta\cal L}[\phi^\prime_k]}{\delta\phi^\prime_j}\right)
\end{eqnarray}
and so if $\phi_i$ undergoes the transformation of Eq. \eqref{eq8} while
\be\label{eq14}
\lambda_i\rightarrow \lambda_i+\lambda_k\frac{\delta H_{ij}}{\delta\phi_k}\xi_j
\ee
then
\be\label{eq15}
S_T = \int dx {\cal L}_T[\phi_k,\lambda_k]
= \int dx \left( {\cal L}[\phi_k] + \lambda_i\frac{\delta{\cal L}[\phi_k]}{\delta\phi_i}\right)
\ee
is left invariant.  

Following the Faddeev-Popov procedure \cite{10,11}, the path integral associated with $S_T$
is supplemented with the factor
\begin{eqnarray}\label{eq16}
  \int{\cal D}\xi_i {\cal D}\zeta_i \delta\left( F_{ij}\,\left(
\left(\begin{array}{c} \phi_j \\ \lambda_j\end{array}\right) +
  \left(\begin{array}{lr} 0 & H_{jk} \\
    H_{jk} &  \lambda_l\frac{\delta H_{jk}}{\delta\phi_l}\end{array}\right) 
  \left(\begin{array}{c} \zeta_k \\ \xi_k\end{array}\right)\right) 
-\left(\begin{array}{c} p_j \\ q_j\end{array}\right)  \right) 
\left|\det F_{ij}\left(\begin{array}{lr} 0 & H_{jk} \\
    H_{jk} &  \lambda_l\frac{\delta H_{jk}}{\delta\phi_l}\end{array}\right)\right|
\end{eqnarray}
as well as
\be\label{eq17}
\int {\cal D}p_i {\cal D}q_i \exp \, i \int dx \left[-\frac{1}{2\alpha}
\left(p_i p_i+2p_i q_i\right)\right]
\ee
if we choose the gauge fixing conditions
\be\label{eq18}
F_{ij} \phi_j = 0 = F_{ij}\lambda_j.
\ee 
Upon exponentiating the determinant in Eq. \eqref{eq16} by using Fermionic ghost
fields, we are left with the generating functional
\begin{eqnarray}\label{eq19}
  Z^2 &=& \int{\cal D}\phi_i{\cal D}\lambda_i{\cal D} \bar c_i{\cal D} c_i 
  {\cal D}\bar\gamma_i{\cal D}\gamma_i 
  \exp\, i\int dx\bigg[
    {\cal L}_T[\phi_k,\lambda_k]
-\frac{1}{2\alpha}(F_{ij}\phi_j)^2-\frac{1}{\alpha}(F_{ij}\phi_j)(F_{ik}\lambda_k) \nonumber \\
&+&\bar c_iF_{ij}\left(H_{jk}+\lambda_l\frac{\delta H_{jk}}{\delta\phi_l}\right)c_k 
+\bar \gamma_i F_{ij} H_{jk} c_k + \bar c_i F_{ij} H_{jk}\gamma_k 
+j_i(\phi_i+\lambda_i) \bigg]
\end{eqnarray}
once we make use of the identity
\be\label{eq20}
\det\left(\begin{array}{lr}
0 & A \\ A & B  
\end{array}\right) =
\det\left(\begin{array}{lc}
0 & A \\ A & A+B
\end{array}\right).
\ee
In the path integral in Eq. \eqref{eq19}, $\lambda_i$, $\gamma_i$, $\bar\gamma_i$ are
Lagrange multipliers associated with the 
equations of motion of the
fields $\phi_i$, $c_i$ and
$\bar c_i$ respectively. It is interesting to note that we also have
\be\label{eq21}
\left|\det\left(\begin{array}{lr}0 & A \\ A & B \end{array}\right)\right|
={\det}^2 A
\ee
and so the effect of the functional integrals over $c_i$, $\bar c_i$,
$\gamma_i$ $\bar\gamma_i$ is to give the square of the one loop
contributions coming from the usual Faddeev-Popov factor in
Eq. \eqref{eq10} (as expected).

This general formalism has been used when considering the Yang-Mills
\cite{8} and Proca \cite{9} model. We now will apply it to the
Einstein-Hilbert action.

\section{The Einstein-Hilbert Action with a Lagrange Multiplier}

We now consider the second order Einstein-Hilbert action
\be\label{eq22}
S_{2EH} = \frac{1}{\kappa^2} \int d^4 x \sqrt{g} R[g_{\mu\nu}]
\;\;\;\; (\kappa^2 \equiv 16\pi G_N)
\ee
The gauge invariance of this action is diffeomorphism invariance. If
$g_{\mu\nu}$ is split into a background metric $\bar g_{\mu\nu}$ and a
quantum field $\phi_{\mu\nu}$ \cite{11}
\be\label{eq23}
g_{\mu\nu}  = \bar g_{\mu\nu} + \kappa \phi_{\mu\nu}
\ee
with indices raised and lowered and covariant differentiation defined
using $\bar g _{\mu\nu}$, then a convenient gauge fixing action 
is
\be\label{eq24}
S_{gf} = - \frac{1}{2\alpha} \int dx \sqrt{\bar g}
\left(\phi^{\mu\nu}_{;\bar\nu} -
\phi^{\nu}_{\nu\, ;}{}^{\bar\mu}\right)^2
\equiv -\frac{1}{2\alpha} \int dx \sqrt{\bar g}
\left[F^{\mu,\alpha\beta}(\bar g) \phi_{\alpha\beta}\right]^2.
\ee 
where ``$;\bar\mu$'' denotes a covariant derivative using the
background metric $\bar g_{\mu\nu}$.

The gauge transformation associated with the action of Eq. \eqref{eq22}
is an infinitesimal coordinate transformation 
\begin{eqnarray}\label{a25}
\delta g_{\mu\nu} &=& \kappa\left[g_{\mu\lambda}\partial_\nu\xi^\lambda
+g_{\nu\lambda}\partial_\mu\xi^\lambda
+\xi^\lambda \partial_\lambda g_{\mu\nu}\right]
\nonumber \\ & = &
\kappa\left[g_{\mu\lambda}\xi^\lambda_{;\nu}
+g_{\nu\lambda}\xi^\lambda_{;\mu}\right]
\end{eqnarray}
and so under Eq. \eqref{eq23}
\begin{eqnarray}\label{a26}
\delta\left(
\bar g_{\mu\nu} + \kappa\phi_{\mu\nu}\right) = \kappa\left[
\bar g_{\mu\lambda} \xi^\lambda_{;\bar\nu} +
\bar g_{\nu\lambda} \xi^\lambda_{;\bar\mu}
+\kappa\left(
\phi_{\mu\lambda}\partial_\nu\xi^\lambda+
\phi_{\nu\lambda}\partial_\mu\xi^\lambda+
\xi^\lambda\partial_\lambda\phi_{\mu\nu}
\right)\right].
\end{eqnarray}
There are two types of gauge transformations associated with
that of Eq. \eqref{a26}. In the first type,
\begin{subequations}\label{a27}
\be\label{a27a}
\delta\bar g_{\mu\nu} =
\kappa\left(\bar g_{\mu\lambda}\xi^\lambda_{;\bar\nu}
+               \bar g_{\nu\lambda}\xi^\lambda_{;\bar\mu}\right)
\ee
\be\label{a27b}
\delta\phi_{\mu\nu} =
\kappa\left(
\phi_{\mu\lambda}\partial_\nu\xi^\lambda+
\phi_{\nu\lambda}\partial_\mu\xi^\lambda+
\xi^\lambda\partial_\lambda\phi_{\mu\nu}
\right)
\ee
\end{subequations}
while in the second type
\begin{subequations}\label{a28}
\be\label{a28a}
\delta \bar g_{\mu\nu} = 0 
\ee
\begin{eqnarray}\label{a28b}
\delta \phi_{\mu\nu} &=&  
\bar g_{\mu\lambda}\xi^\lambda_{;\bar\nu}+
\bar g_{\nu\lambda}\xi^\lambda_{;\bar\mu}+\kappa\left(
\phi_{\mu\lambda}\partial_\nu\xi^\lambda+
\phi_{\nu\lambda}\partial_\mu\xi^\lambda+
\xi^\lambda\partial_\lambda\phi_{\mu\nu}
\right) \nonumber \\ &\equiv&
H_{\mu\nu , \lambda}(\phi)\xi^\lambda.
\end{eqnarray}
\end{subequations}
The gauge fixing of Eq. \eqref{eq24} does not break the gauge
invariance of Eqs. \eqref{a27}, but breaks that of
Eq. \eqref{a28}. By use of Eqs. \eqref{eq9} and \eqref{eq10} we can
find the gauge fixing and Faddeev-Popov ghost Lagrangians that follow
from Eqs. \eqref{eq24} and \eqref{a28} for $S_{2EH}$ of
Eq. \eqref{eq22} alone.

If the background metric is flat (ie $\bar g_{\mu\nu}
=\delta_{\mu\nu}$) then
\be\label{a29}
S_{gf} = -\frac{1}{2\alpha}\int d^4 x\left(
\phi_{\mu\nu,\nu} - \phi_{\nu\nu ,\mu}
\right)^2
\ee
and \cite{10,11}
\be\label{a30}
S_{FP} = \int d^4 x \bar c_\mu\left(\partial^2\delta_{\mu\nu}+
\kappa M_{\mu\nu}(\phi)\right)c_\nu
\ee
where
\be\label{a31}
M_{\mu\nu}(\phi) = \overset{\leftarrow}{\partial_\sigma}
\left(\phi_{\mu\sigma ,\nu} +\phi_{\nu\sigma ,\mu}
\right)  
-\frac 1 2
\overset{\leftarrow}{\partial_\mu}
\left(\phi_{\lambda\lambda,\nu} + \phi_{\lambda\nu,\lambda}\right).
\ee

Variation of $g_{\mu\nu}$  in Eq. \eqref{eq22}  leads to \cite{12}
\be\label{eq27}
\delta S_{2EH} =  -\frac{1}{\kappa^2}\int dx \delta g_{\mu\nu}\sqrt{g} \,G^{\mu\nu}(g)\;\;\;
\left(G^{\mu\nu}\equiv R^{\mu\nu} - \frac 1 2 g^{\mu\nu} R\right).
\ee

We will now adapt the arguments of the preceding section to deal with 
background field quantization of the Einstein-Hilbert action when
using a Lagrange multiplier field to suppress higher loop
contributions to
the effective action. The Lagrange multiplier field
$\lambda_{\mu\nu}$ associated with the metric $g_{\mu\nu}$ has a
background part $\bar\lambda_{\mu\nu}$ and a quantum part
$\psi_{\mu\nu}$ 
\be\label{b33}
\lambda_{\mu\nu} = \bar\lambda_{\mu\nu} + \kappa\psi_{\mu\nu}.
\ee
We consider the action (much like that in Eq. \eqref{eq3})
\be\label{b34}
S_T = \int d^4 x {\cal L}_T =
\frac{1}{\kappa^2}\int d^4 x\sqrt{\bar g + \kappa\phi}\left[
R(\bar g+\kappa\phi) - \left(\bar\lambda^{\mu\nu}+
\kappa\psi^{\mu\nu}\right) G_{\mu\nu}(\bar g +\kappa\phi)
\right].
\ee
Eq. \eqref{b34} follows from Eq. \eqref{eq22} just as Eq. \eqref{eq3}
follows from Eq. \eqref{eq1}.

If $\phi_{\mu\nu}$ undergoes the transformation of Eq. \eqref{a28b},
then using the arguments leading to Eqs. (\ref{eq12}, \ref{eq14}) we
see that
\begin{subequations}\label{b35}
\be\label{b35a}
\delta\psi_{\mu\nu} = H_{\mu\nu,\lambda} \zeta^\lambda
\ee
and
\be\label{b35b}
\delta\lambda_{\mu\nu} = \frac{1}{\kappa}\left(
\bar\lambda_{\alpha\beta} + \kappa\psi_{\alpha\beta}\right)
\frac{\delta H_{\mu\nu,\lambda}}{\delta\phi_{\alpha\beta}}\xi^\lambda
\ee
\end{subequations}
are gauge transformations associated with $\psi_{\mu\nu}$ 
(with $\delta\bar\lambda_{\mu\nu} = 0$).

Next we insert into the path integral associated with quantizing $S_T$
a factor of unity much like that of Eq. \eqref{eq16} 
\begin{eqnarray}\label{b36}
&  \int{\cal D}\xi_\mu {\cal D}\zeta_\mu 
\delta\left( F^{\mu,\alpha\beta}\,\left(
\left(\begin{array}{c} \phi_{\alpha\beta} \\ \psi_{\alpha\beta}\end{array}\right) +  \left(\begin{array}{cc} 0 & H_{\alpha\beta,\rho} \\
    H_{\alpha\beta,\rho} &               \frac{1}{\kappa}\left(\bar\lambda_{\pi\tau}
+\kappa\psi_{\pi\tau}\right) 
\frac{\delta H_{\alpha\beta,\rho}}{\delta\phi_{\pi\tau}}
\end{array}\right) 
  \left(\begin{array}{c} \zeta_\rho \\ \xi_\rho\end{array}\right)\right) 
-\left(\begin{array}{c} p^\mu \\ q^\mu\end{array}\right)  \right) 
\nonumber \\ &
\left|\det F^{\mu,\alpha\beta}\left(\begin{array}{cc} 0 & H_{\alpha\beta,\rho} \\
    H_{\alpha\beta,\rho} &              \frac{1}{\kappa}\left(\bar\lambda_{\pi\tau}
+\kappa\psi_{\pi\tau}\right) 
\frac{\delta H_{\alpha\beta,\rho}}{\delta\phi_{\pi\tau}}
\end{array}\right)\right| .
\end{eqnarray}
In addition, we insert a constant
\be
\int{\cal D} p^\mu {\cal D} q^\mu\exp -\frac{1}{2\alpha}\int d^4 x 
\sqrt{\bar g}\left(p^\mu p_\mu+ 2 p^\mu q_\mu\right)
\ee
so that much like Eq. \eqref{eq19} we have the generating functional
\begin{eqnarray}\label{b38}
  Z^2 &=& \int{\cal D}\phi_{\mu\nu}{\cal D}\psi_{\mu\nu}
{\cal D} \bar c_\mu{\cal D} c_\mu 
  {\cal D}\bar\gamma_\mu{\cal D}\gamma_\mu 
  \exp\, -\int d^4x\bigg\{
    {\cal L}_T + \sqrt{\bar g}\bigg [
-\frac{1}{2\alpha}\left(
(F^{\mu,\alpha\beta}\phi_{\alpha\beta})^2
+2(F^{\mu,\alpha\beta} \phi_{\alpha\beta})
   (F_{\mu,\gamma\delta} \psi^{\gamma\delta})\right)
\nonumber \\
&+&\bar c_\mu F^{\mu,\alpha\beta}\left(H_{\alpha\beta,\nu}
+\frac{1}{\kappa}\left(\bar \lambda_{\pi\tau}
+\kappa\psi_{\pi\tau}\right)
\frac{\delta H_{\alpha\beta,\nu}}{\delta\phi_{\pi\tau}}\right)c^\nu 
\nonumber \\
&+&
\bar c_\mu F^{\mu,\alpha\beta} H_{\alpha\beta,\nu} \gamma^\nu+
\bar \gamma_\mu F^{\mu,\alpha\beta} H_{\alpha\beta,\nu} c^\nu\bigg ]
+\sqrt{\bar g}(\phi_{\mu\nu}+\psi_{\mu\nu}) j^{\mu\nu}\bigg\}.
\end{eqnarray}
Again using Eq. \eqref{eq21}, we see that the ghost contribution to
Eq. \eqref{b38} is $\det^2(F^{\mu,\alpha\beta} H_{\alpha\beta,\nu})$
which is the square of the Faddeev-Popov contribution arising when
considering the Einstein-Hilbert action alone.
The Lagrange multiplier fields associated with the quantum fields 
$\phi_{\mu\nu}$, $c_\mu$ and $d_\mu$ are $\psi_{\mu\nu}$, $\xi_\mu$ and $\zeta_\mu$ respectively.

We now can make the usual choice of background field 
metric $\bar g_{\mu\nu} = \delta_{\mu\nu}$ (flat space).
Since the vertices containing $N$ external fields 
$(\phi_{\mu\nu},\psi_{\mu\nu})$ can be obtained 
from Eq. \eqref{b38} by expanding $\sqrt{g}G^{\mu\nu}$  up to ${\cal O}(\phi^{N-1})$
in $\phi_{\mu\nu}$,
the derivation of the Feynman rules 
needed for calculations of one loop Green's functions following 
from Eq. \eqref{b38}
is simpler than in the usual approach following just from $S_{2EH}$.
Using this approach we have generated the vertices up to the four
external fields $\phi_{\mu\nu}$ and explicitly
verified that $\langle \phi\psi\rangle =   \langle\phi\phi\rangle$, 
$\langle \phi\phi\psi\rangle =   \langle\phi\phi\phi\rangle$ and  
$\langle \phi\phi\phi\psi\rangle =  \langle\phi\phi\phi\phi\rangle$ in agreement
with general expressions which follow from Eq. \eqref{eq7}.
This, together with the fact that the combinatorial factors of loop diagrams with mixed propagators 
are twice the ones in the usual theory and 
also that there are two ghost fields in Eq. \eqref{b38}, 
is sufficient to  demonstrate that the results for all the one-loop 
diagrams will be twice the corresponding results in the usual formulation 
of quantum gravity. 
We note that vertices in the 2EH action become simpler if we were to
use the first order (Palatini) action 1EH \cite{13,14}.

When using background field quantization then
both dimensional arguments and explicit calculation show that all the
one loop divergences for the 2EH effective 
action alone are of the form \cite{1,2} 
\be\label{eq29}
\int dx \left[\sigma_1\ R^2 +\sigma_2 R_{\mu\nu}^2 +  \sigma_3 R_{\mu\nu\lambda\sigma}^2\right].
\ee
By the Gauss-Bonnet theorem, 
$R^2 - 4 R^2_{\mu\nu} 
+ R^2_{\mu\nu\rho\sigma}$ $=$ (surface term),
the expression in the bracket is a surface term, so that
$R_{\mu\nu\lambda\sigma}^2$ can be
expressed in terms of $R^2$ and $R_{\mu\nu}^2$ and the one loop
divergences 
in $n=4-\epsilon$ dimensions ($\sqrt{\bar g}/(8\pi^2\epsilon)(\frac{1}{120} 
\bar R^2+\frac{7}{20}\bar R^2_{\mu\nu})$)
are proportional to terms that vanish when the equations
of motion are satisfied.
This means that they can be removed by a field redefinition \cite{1,2} 
when working with the $S_{2EH}$ alone
or by rescaling the field $\lambda_{\mu\nu}$ in Eq. \eqref{b38}. The
two loop divergences that arise using $S_{2EH}$ alone can only be
removed if a new term appears in the classical action; introduction of the
Lagrange multiplier field circumvents
this problem. This approach preserves the structure of the
conventional theory to one-loop order but suppresses higher loop
contributions where non renormalizable (by power counting) divergences
arise. In this way, pure gravity effectively becomes renormalizable.

There are differences between the divergences of Eq. \eqref{eq29}
arising when examining 
one-loop corrections to the Einstein-Hilbert action,
and those appearing in the Yang-Mills theory. With Yang-Mills theory and
the Dyson procedure, the fields and couplings appearing in the
original classical action can be rescaled in order to absorb
divergences as all divergences appear in terms that are of the same
functional form as ones in the original classical action. This feature
is also present when the Yang-Mills action is supplemented by a
Lagrange multiplier term that eliminates higher loop corrections \cite{8}.
However, when considering the Einstein-Hilbert action alone where
divergences at one-loop order are of the form of Eq. \eqref{eq29}, this
is no longer the case as neither $R^2$, $R_{\mu\nu} R^{\mu\nu}$ nor
$R_{\mu\nu\rho\sigma} R^{\mu\nu\rho\sigma}$ appear in the action of
Eq. \eqref{eq22}. But since Eq. \eqref{eq29} is proportional to
$G_{\mu\nu}$ (once the Gauss-Bonnet theorem is used), these divergences
can be absorbed by shifting the metric as its equation of motion is
$G_{\mu\nu}=0$ (from Eq. \eqref{eq27}). The introduction of the
Lagrange multiplier field makes it possible to eliminate divergences
by an alternate shift; instead of shifting the metric as in refs. \cite{1,2},
it is now possible to shift the Lagrange multiplier field in order to
absorb one-loop divergences. This will prove possible even when
matter fields are present.

In more detail, we see that from Eq. \eqref{b38} we have the
contribution
\be\label{c40}
S_{\bar\lambda} = -\frac{1}{\kappa^2}\int d^4 x\sqrt{\bar g}
\bar\lambda^{\mu\nu} {\bar G}_{\mu\nu}
\ee
in the effective action, as well as the divergent piece
\begin{subequations}
\be\label{c41a}
S_{div} = \frac{2}{8\pi^2\epsilon}\int d^4 x\sqrt{\bar g}
\left(\frac{1}{120}{\bar R}^2+\frac{7}{20}{\bar R_{\mu\nu}}^2\right). 
\ee
From the definition of $G_{\mu\nu}$ in Eq. \eqref{eq32}, this becomes
\be\label{c41}
= \frac{1}{4\pi^2\epsilon}\int d^4 x\sqrt{\bar g}
\left(\frac{7}{20} \bar G^{\mu\nu}
+\frac{1}{120}\bar G \bar g^{\mu\nu}\right)\bar G_{\mu\nu}. 
\ee
\end{subequations}
The divergence in the effective action can be removed by shifting
$\bar\lambda_{\mu\nu}$ to $\bar\lambda^R_{\mu\nu}$  where
\be\label{c42}
\bar\lambda^R_{\mu\nu} = \bar\lambda_{\mu\nu}
- \frac{ \kappa^2}{4\pi^2\epsilon}
\left(\frac{7}{20} \bar G^{\mu\nu}
+\frac{1}{120}\bar G \bar g^{\mu\nu}\right). 
\ee
There is no need to renormalize $\bar g_{\mu\nu}$ or $\kappa^2$ 
and no further divergences can arise since no radiative effects occur
beyond one-loop order.

The mass dimension of $g_{\mu\nu}$ is $[\mbox{mass}]^0$ in $n$ spatial
dimensions, and $R$, $R_{\mu\nu}$, $G$ and $G_{\mu\nu}$ are 
$[\mbox{mass}]^2$. 
Consequently, $\kappa^2$ and $\lambda_{\mu\nu}$
have mass dimensions $[\mbox{mass}]^{\epsilon-2}$ and
$[\mbox{mass}]^{0}$. As a result, in $n=4-\epsilon$ dimensions,
$\kappa^2$ incorporates an arbitrary mass parameter $\mu^2$ so that
\be\label{c43}
\kappa^2 = 16\pi G_N \mu^\epsilon.
\ee
From Eqs. (\ref{c41},\ref{c42}) we see that as $\epsilon\rightarrow 0$,
the effective action will contain an arbitrary term proportional to
$\ln\mu$ once the renormalization of Eq. \eqref{c42} is
taken into account. This arbitrariness is compensated by an
arbitrariness in $\bar\lambda^R_{\mu\nu}$ ($\bar g_{\mu\nu}$ and
$\kappa^2$ are not altered by changes in $\mu$). We see that together
Eqs. (\ref{c42}, \ref{c43}) result in 
\be\label{c44}
\bar\lambda^R_{\mu\nu} = 
- \frac{4}{\pi} G_N
\left(\frac{7}{20} \bar G^{\mu\nu}
+\frac{1}{120}\bar G \bar g^{\mu\nu}\right) 
\ln\left(\frac{\mu}{\Lambda}\right) .
\ee
where $\mu/\Lambda$ is fixed by experiment. Together Eqs.
(\ref{c40}, \ref{c41}, \ref{c42}, \ref{c44}) imply that 
$S_{\bar\lambda}+ S_{div}$ give a contribution to the effective
action of
\be\label{c45}
S_{new} = \frac{1}{4\pi^2}\ln\left(\frac{\mu}{\Lambda}\right) 
\int d^4 x \sqrt{\bar g}\left(
\frac{1}{120} {\bar R}^2 + \frac{7}{20} \bar R_{\mu\nu}
\bar R^{\mu\nu}
\right).
\ee
The consequence of renormalization is thus to give to the effective
action a contribution quadratic in $\bar R$ and $\bar R_{\mu\nu}$ with
undetermined strength  
$\frac{1}{4\pi^2}\ln\left(\frac{\mu}{\Lambda}\right)$. 

We now examine how the Lagrange multiplier field can be used when the
metric couples to a self interacting scalar field so that
renormalizability is retained while all higher loop contributions
involving internal scalar lines still contribute to the effective
action. Such higher loop contributions must be included if our
approach were to be applied to a fully covariant version of the
Standard Model that is consistent with experiments.

\section{A covariant action with a self interacting scalar field}
The action we will consider is of the form $S=S_{2EH} + S_\lambda + S_\phi$
where
\be\label{eq30}
S_{2EH} = \frac{1}{\kappa^2}\int d^4x\sqrt{g} R(g_{\mu\nu}) 
\ee
\be\label{eq31}
S_{\lambda} = -\frac{1}{\kappa^2}\int d^4x\sqrt{g} \lambda_{\mu\nu}
G^{\mu\nu}(g_{\mu\nu}) 
\ee
\be\label{eq32}
S_{\phi} = \int d^4x\sqrt{g} \left(
\frac 1 2 g^{\mu\nu}\partial_\mu\phi \partial_\nu\phi
-\frac 1 2(m^2-\kappa R)\phi^2 -\frac{1}{4!}\lambda\phi^4+\Lambda
\right).
\ee
The contribution $S_{2EH}+S_\lambda$ by itself was examined in the
preceding section; there it was shown that the presence of the field
$\lambda_{\mu\nu}$ restricts the radiative corrections to one loop
  order and these one-loop corrections are twice those arising from
  $S_{2EH}$ alone. Adding $S_\phi$ does not change this
  conclusion. The new diagrams arising upon including $S_\phi$ all have
  internal propagators arising from the scalar field $\phi$ and either
  the metric or the scalar field on the external legs. Those diagrams
  with internal lines coming from the scalar propagator occur at
  arbitrarily high order in the loop expansion.

The divergences that arise due to quantum effects in the model of a
scalar field in the presence of a background metric are analyzed in
ref. \cite{15} (see ch. 3 and references there in). It is shown that
divergences can either be absorbed by renormalizing the parameters and
fields occurring in $S_\phi$ itself, or arise due to vacuum effects
when the background space-time is curved. In this later case, the
divergences can be absorbed either by renormalizing $\sqrt{g}R$ in
$S_{2EH}$, by renormalizing the cosmological constant term
$\sqrt{g}\Lambda$ in $S_\phi$, or are of the  form of Eq. \eqref{eq29}
in which case they can be absorbed by renormalizing the Lagrange
multiplier field $\lambda_{\mu\nu}$ as are the divergences arising
from $S_{2EH} + S_\lambda$ which are
discussed in the preceding section. It
thus proves possible to eliminate all divergences arising from 
$S_{2EH} + S_\lambda + S_\phi$.

The same conclusion can be reached if in addition to scalar fields,
there are also spinor and vector fields present. As a result, it
should be possible, using the Lagrange multiplier field, to have a
renormalizable Standard Model that is fully covariant. It should also
be possible to use a Lagrange multiplier to eliminate radiative
effects beyond one-loop order in Supergravity Models.

There are some interesting consequences to having introduced this
Lagrange multiplier. Its presence eliminates all higher loop effects
(which are known to give rise to non-renormalizable divergences),
while at the same time leaves the theory unitary. The classical
consequences of the Einstein-Hilbert action are all retained. Matter
fields can be coupled to the metric without affecting these features,
and the matter fields themselves are coupled only to a background
metric (which may have consequences when considering Hawking
radiation). One-loop correction to the Einstein-Hilbert action are all
in principle computable. Just as introducing the Higgs makes quantizing
the Standard Model viable, we may consider the Lagrange multiplier as
a possible candidate for a mechanism to reconcile gravity and the
quantum theory.


\acknowledgments 

F. T. B. and J. F. would like to thank CNPq (Brazil) for a grant. 
D. G. C. M. would like to thank T. N. Sherry for collaboration at the
early stages of this work, 
Roger Macleod for a  suggestion, 
Fapesp (Brazil) for financial support (grant number 2018/01073-5) 
and Universidade de S\~ao Paulo for its warm hospitality.



\end{document}